**TITLE PAGE**

**Title:**

Neurofeminism: feminist critiques of research on sex/gender differences in the neurosciences

**Authors:**

Kassandra Friedrichs[1], Philipp Kellmeyer[2]*

1 – University College Freiburg, Albert-Ludwigs-University of Freiburg, Bertoldstr. 17, DE-79098 Freiburg i. Br. (Germany)

2 – Human-Technology Interaction Lab, University Medical Center Freiburg, Department of Neurosurgery, Breisacher Str. 64, DE-79106 Freiburg i. Br.

Phone: +49 761 203 97446

ORCID: 0000-0001-5538-373X (Philipp Kellmeyer)

* corresponding author

**Acknowledgments and funding:**

The work of author P.K. was partially supported by a grant (00.001.2019) of the Klaus Tschira Foundation, Germany.

**Running title**

Sex/gender difference brain research

**Word count:** 7481 (excl. title, abstract, references), 9870 (incl. title, abstract, references)

**Figure count:** 0

**Abstract**

Over the last three decades, the human brain, and its role in determining behavior have been receiving a growing amount of attention in academia as well as in society more generally. Neuroscientific explanations of human behavior or other phenomena are often especially appealing to lay people. Therefore, neuroscientific explanations that can affect individuals, groups, or social relations in general should be formulated in a careful and responsible way. One field in which especially feminist scholars request more caution is the neuroscientific examination of sex/gender differences. Feminist scholars have described various ways in which sexist bias might be present in neuroscientific research on sex/gender differences. In this context, they coined the term 'neurosexism' to describe the entanglement between neuroscientific work and sexist ideology, and 'neurofeminism' as a response to that. Here, we aim to give an overview over the contemporary neurofeminist literature. In the first part, common levels of analysis in the neurofeminist literature are presented and the research level is explored in more detail. In the second part, some common developments in more recent neurofeminist scholarship are discussed. For this, we review recent publications with the aim to provide neuroscientists with a solid understanding of neurofeminist criticism so that they may evaluate neuroscientific claims about on sex/gender differences from this critical perspective.

**Introduction**

 In 1990, the sitting U.S. president George H. W. Bush signed a proclamation that declared the following decade as the "Decade of the Brain" (Roy, 2016a). The goal of this proclamation was to stress the importance of brain research and to foster collaborations between federal agencies and private companies for joint brain research programs. In this context, the first "Human Brain Project" ("The Human Brain Project," 2021), a multi-center research program to promote the development of progressive software for innovative brain research, was launched. It was followed up by other largescale brain research projects, such as the U.S. "BRAIN Initiative" ("Brain Initiative," 2021), the EU-based "Human Brain Project" ("The Human Brain Project," 2021) and the Swiss "Blue Brain Project" ("Blue Brain," 2021). According to Schmitz & Höppner (2014a), these programs appear to promise to eventually provide an "all-explaining knowledge framework with which to explain 'the human'" (p. 6).

This exciting promise from brain research has captivated many researchers in the neurosciences and computer sciences and led to a surge in the popularization of neuroscience research in the media and in society more generally. As a result, neuroscience-based explanations of human behavior and mental characteristics enjoy a special epistemic authority (Weisberg *et al.*, 2008; Fine, 2012). This means that lay people find explanations about psychological phenomena more convincing when they contain neuroscientific information, even if that information is not logically relevant to explaining the appropriate phenomena. This is illustrated by the abundance of references to neuroscientific findings or applications thereof in newspapers and the great amount of popular neuroscience literature (Weisberg *et al.*, 2008). Weisberg et al. (2008) describe this public interest in neuroscientific facts and explanations and the tendency to view explanations as more satisfying when they contain neuroscientific information as the "seductive allure" (p. 470) of neurosciences. The authors highlight that this phenomenon entails a certain responsibility and caution when using neuroscientific facts to support a claim, as these might influence the opinion of people more than what would be desired or reasonable.

This seductive allure of theories based on neuroscientific findings can differ for the type of phenomenon to be explained. Especially (pseudo-)neuroscientific theories that aim to explain purported behavioral and mental differences between women and men receive increased attention. Countless popular neuroscience books on this topic have been published. Even though some of these books have been criticized for being stereotype-laden and using the neuroscientific literature in a careless, misleading, or even clearly incorrect manner, these lay books enjoy quite an extensive popularity (Fine, 2008).[1]

This fascination with SGDs in popular neuroscience books is not arbitrary. While numerous neuroscientists nowadays actively distance themselves from the bold claims of these books and their often-inadequate use of neuroscientific findings, the neurosciences themselves have a long tradition of trying to demonstrate how the brains of women and men are different and that these differences are predominantly in accordance with common sex/gender stereotypes. For example, during the latter half of the 19[th] century, neuroscientists vigorously tried to demonstrate that the brains of women were smaller than those of men, and that, as a result, women were less intelligent than men (Kaiser *et al.*, 2009; Bluhm, 2012). This practice of using neuroscientific research and its results to promote sexist ideas has been termed "neurosexism" (Fine, 2008; see also Bluhm, 2012; Fine, 2013; Hoffman & Bluhm, 2016). Or as Fine (2008) phrases it, neurosexism can be defined as "the ugly rush to cloak old-fashioned sexism in the respectable and authoritative language of neuroscience" (p. 69). Feminist scholars from various disciplines have carefully examined possible sexist tendencies in the neurosciences, from their predecessors like phrenology, to modern-day neuroscientific research. They have argued that even contemporary neuroscience research is influenced by and reproduces sexist biases in various ways. The work aimed at critically analyzing and responding to this entanglement between sexist ideology and neuroscience can be subsumed under the term neurofeminism, a phrase which Bluhm et al. (2012) first popularized. Due to the seductive allure of neuroscientific 'facts', neurofeminist work is of great importance to limit the dissemination and uncritical acceptance of neurosexist claims.

During the late 1970s, scholars began to critically examine the foundations of modern science from a feminist perspective (Leavitt & Gordon, 1988; Rosser, 1989; Harding, 1991; Roy, 2016b). Drawing on approaches that were established in the social sciences, humanities, and the women's movement, these scholars developed a discipline now referred to as feminist science studies. The central effort of feminist science studies scholars has been to "carefully explore the myriad ways in which sexist biases [affect] the nature and practice of science" (Tuana, 1989). According to Bluhm et al. (2012), feminist neuroscientists were among the first to analyze their own field's theories and practices of knowledge production through a feminist lens. Ruth Bleier was one of these revolutionary scholars and contributed significantly to a critical examination of the foundations of modern biology and neuroscience. According to her, "the most publicized arena for demonstrating women's inferiority" (Leavitt & Gordon, 1988) during the 1980s was her own area of research, the neurosciences. As a result, Bleier and other feminist scholars examined how the neurosciences and its preceding and associated disciplines have been instrumentalized to demonstrate that women are inferior to men.

---

[1] For example, Louann Brizendine's book *The Female Brain* (Brizendine, 2007) is a "New York Times Bestseller", has been translated into 21 languages and newspapers, magazines and TV shows around the world discussed it. This can be seen as evidence for a widespread desire to be able to explain observed or purported sex/gender differences (SGDs) in behavior, performance of certain tasks, and mental states (Fine, 2008).

Today, neuroscientific research on SGDs mainly employs functional neuroimaging techniques to measure brain function and not just structure. The main techniques used are positron emission tomography (PET) and functional magnetic resonance imaging (fMRI).[1]

This paper aims to provide an overview over neurofeminist scholarship produced over the last two decades.

**Methodological approach**

In the following, we aim to provide an overview of contemporary neurofeminist critiques of the research on SGDs in the neurosciences, specifically using functional neuroimaging techniques. Firstly, we aim at illustrating common levels of criticism and outlining the neurofeminist argumentations on one of these levels, with its three sublevels, in more detail. Secondly, we intend to show how this criticism has developed over the last two decades. For our analysis of neurofeminist critiques in these two parts, we chose to focus on publications from the NeuroGenderings Network. This is motivated by the fact that neurofeminist voices appear in several disciplines and contexts, and therefore, it is difficult to fully map the literature. For instance, the papers by Joel et al. (2015) and Alon et al. (2020) which we will later identify as important contributions to feminist neuroscience, do not include any mention of neurofeminism or even 'feminist critique' in their texts. As a result, an online search with the keywords 'neurofeminism' or 'feminist criticism' would be unable to detect these articles, despite the authors being members of the NeuroGenderings Network and their works are used to further neurofeminist goals. As the NeuroGenderings Network can be seen as an important agent in the realm of neurofeminist critique due to its structuring, systemizing and creative role, we chose the work of the network as the focus. We additionally confined the analysis to work that is directly or indirectly relevant to the research on SGDs in the neurosciences, especially concerning studies using neuroimaging techniques, as this kind of research has become increasingly popular and influential during the last two decades (Kaiser *et al.*, 2009).

---

[1] One prominent actor in neurofeminist discussions and approaches today is the NeuroGenderings Network. It was formed at a conference titled 'NeuroGenderings: Critical Studies of the Sexed Brain' at Uppsala University in March 2010 ("The NeuroGenderings Network," n.d.).

For the first part of our analysis, the identification of common levels of neurofeminist criticism, we mainly focus on the two anthologies published by the network: 'Neurofeminism: Issues at the Intersection of Feminist Theory and Cognitive Science', published in 2012, and 'Gendered Neurocultures: Feminist and Queer Perspectives on Current Brain Discourses', published 2014. Apart from the articles in the anthologies, we also incorporated frequently referenced articles into our analysis. Initially, to provide a general overview over the different levels, we focus on a review by Schmitz & Höppner (2014a). The review is a modified version of their introduction to the latter anthology (Schmitz & Höppner, 2014b) and nicely outlines the different themes, directions, questions, problems, and goals that underly the work of the NeuroGenderings Network and its members. After having established an overview over the general levels of neurofeminist work, we identify three levels that are especially common in neurofeminist discourses. These levels are: 1) conceptual criticism, 2) methodological criticism, and 3) criticism about theory building. We outline the most dominant neurofeminist argumentations for each of these levels.

In the second part of the paper, we provide an analysis of how the neurofeminist critique has changed during the last years by reviewing articles of members of the NeuroGenderings Network published after the review by Schmitz & Höppner (2014a). We chose the publication of the review as a 'cutoff-point' because the review is, to our knowledge, the first systematic overview of neurofeminist work conducted by the NeuroGenderings Network. Therefore, reviewing and analyzing the literature published after the review is an attempt to update and expand on the review. We identified the articles included in our analysis from the website of the NeuroGenderings Network ("Publications," 2014) and screened them for relevance to our paper. More specifically, we only included papers that demonstrate some relevance to neuroscientific research on SGDs. Furthermore, we especially focused on those publications that seemed to relate to the three levels that we analyzed in detail in the first part of our paper.

**Results**

*A multilevel approach to classifying neurofeminism scholarship*

For classifying different levels of criticism in neurosexism and neurofeminism research, we follow a multilevel approach similar to the one proposed by Schmitz & Höppner (2014a).

Firstly, according to Schmitz & Höppner (2014a), neurofeminist scholars reflect how assumptions and norms about gender and intersected categories enter contemporary brain research (*research level*). This can occur on several sublevels: A) on the conceptual level, where feminist scholars critically examine the definitions of sex/gender used. B) on the methodological level, where they highlight the inconsistencies of findings and methodological issues. And c), on the level of theory building, where neurofeminists critically assess underlying ideologies, such as biological determinism. Secondly, scholars also discuss the impact this research has on the gendered social order and gendered culture (*sociocultural level*). In the discussion of the impacts of neuroscientific research on society, particular attention is devoted to the role of popular science publications and media. For example, feminist scholars analyze how scientific findings are referenced in popular media to legitimize social hierarchies. Another part of neurofeminist work is the development of approaches for a "more gender adequate neuroscientific research" (Schmitz & Höppner, 2014a). Schmitz & Höppner (2014a) refer to this kind of work as "feminist neurosciences" (p. 2). Lastly, neurofeminist scholars also consider the uses and misuses of their own discussions and concepts (*sociopolitical level*). For instance, while the plasticity paradigm is regarded as a highly valuable concept for overcoming neurodeterminism by some feminist scholars, others point to its possible problematic consequences, such as using it for neuro-enhancement in a neoliberal fashion (Schmitz & Höppner, 2014a).

The work on these levels is of course often closely related and intersects. For instance, feminist scholars have observed how certain methodological features are due to a belief in a strong sex/gender essentialism (*research level*) and how these methodologies then further reinforce the belief in sex/gender essentialism in the public (*sociocultural level*). As a consequence, feminist scholars call for a change of these methodological practices (in the sense of 'feminist neuroscience').

In the following, given constraints of scope and space, we mostly address the *research level* of neurofeminist work and its three sublevels (conceptual, methodological, theory-building).

*Research Level I: Conceptual criticism*

Feminist scholars of the NeuroGenderings Network have argued that stereotypical preconceptions about sex/gender can influence research already at the level of study design, especially with the choice of categories to be compared. Feminist scholars criticize that the operationalization of the variable 'sex/gender' and interpretations about it are characterized by a strong commitment to sex dimorphism in most research on SGDs (Dussauge & Kaiser, 2012; Jordan-Young & Rumiati, 2012; Meynell, 2012; Fine, 2013; Fine *et al.*, 2013; Joel, 2014; Kaiser, 2014). Sex dimorphism describes the idea that sex/gender presents itself in two distinct forms, without overlap between those forms. Feminist scholars argue that a general sex dimorphism applies neither to human behavior, nor to brain structure and function – even though SGDs in behavior and brain structure/function are found in neuroscientific studies. Furthermore, these scholars point out the consequences that an underlying commitment to sex dimorphism can have on the results and interpretations of research on SGDs.

According to Joel (2014), a sex dimorphic view of the brain and behavior follows from the incorrect assumption that the characteristics that underlie 3G-sex, which defines sex in terms of a person's genetic, gonadal, and genital make-up, also hold true for other domains. The majority of people fall into one of two categories in each aspect of 3G-sex. Additionally, this system of categorization exhibits a high degree of "internal consistency" (Joel, 2014), meaning that a person qualifying as 'female' in one aspect of 3G-sex likely also exhibits the female form of another feature of 3G-sex. Therefore, 3G-sex can be viewed as highly dimorphic. These characteristics of 3G-sex are, however, not applicable to gendered behavior and the brain, as several authors argue (Jordan-Young & Rumiati, 2012; Joel, 2014). These authors point out that SGDs in brain activation patterns can only be found at the population level, are rather distributed along a continuum, and show a considerable amount of overlap. Accordingly, when comparing the brain activation patterns of several women and men during a task X, women might display a significantly different activation in a certain brain area Y. However, this does not mean that *all* women would exhibit a similar 'female' activation in brain area Y during task X, but rather, many women might actually show a 'male' or an intermediate activation pattern (Joel, 2014). Additionally, Joel (2014) discusses evidence from animal research against the idea that SGDs in the brain are highly internally consistent. Accordingly, neurofeminists argue that claims presenting SGDs in brain activation in a sexually dimorphic way, such as assertions about 'female' and 'male' brains found in popular science books and the media, are misleading. In addition, these claims perpetuate sociocultural preconceptions that then help to normalize and justify further neuroscience research on sex/gender dimorphism (Jordan-Young & Rumiati, 2012; Meynell, 2012; Joel, 2014).

Despite the evidence against brains being sexually dimorphic, feminist scholars have outlined how a substantial amount of research on SGDs still presupposes this sex dimorphism in brains and behavior (Dussauge & Kaiser, 2012; Meynell, 2012; Fine, 2013; Kaiser, 2014). Some authors point to, for instance, certain methodological constraints as evidence for a strong underlying conceptual assumption of sex dimorphism. Meynell (2012) argues that the only way that small sample sizes could be justified in SGDs research (which is a common phenomenon, see later discussion) is "if you are already committed to a strong sex dimorphism" (p.25).

*Research Level II: Methodological criticism*

Feminist scholars have argued that neuroimaging research on SGDs tends to have various methodological shortcomings that result in sexist biases in research, e.g., by emphasizing neurobiological differences rather than similarities (Bluhm, 2012; Dussauge & Kaiser, 2012; Fine, 2012, 2013; Grossi & Fine, 2012; Hoffman, 2012; Jordan-Young & Rumiati, 2012; Meynell, 2012; Roy, 2012). These sexist biases can become apparent in each phase of experimentation, including data gathering, analysis and interpretation, and presentation. According to Hoffman & Bluhm (Hoffman & Bluhm, 2016) some methodological issues are so severe as to constitute a "violation of accepted scientific practice" (p. 721). Of course, 'bad science' can affect any area of research. However, neurofeminists have argued that some instances of neurosexism they describe are not just 'bad science', as the methodological issues primarily seem to support one type of reasoning: the identification and interpretation of SGDs in accordance with existing stereotypes (Fine, 2013). In the following, we summarize the methodological issues most commonly outlined by feminist scholars on each level of research.

*Research Level IIa: Criticism of research design, data collection and analysis practices*

Concerning research design, one important aspect of neuroimaging study design that has been highlighted in recent years is the consideration of statistical power. Scientific research on differences between groups requires a large number of participants so that any random variation within the population is "wash[ed] out" (Meynell, 2012). This reduces the chance of committing a false-positive error, so finding a difference between the two groups where no difference actually exists. Button et al. (2013) also refers to this issue as "power failure" (p. 365), which is of substantial concern in cognitive neuroscience. According to Wallentin (2009), neuroimaging studies are especially vulnerable to being underpowered because an adequate number of participants is needed to balance out nuisance variables that affect the imaging signal. Therefore, Thirion et al. (2007) recommend a sample size of at least 20 participants per group. Yet, several scholars have shown that the number of participants in studies on SGDs is often lower than that (Kaiser *et al.*, 2009; Fine, 2010, 2013; Meynell, 2012; Rippon *et al.*, 2014). For example, an analysis by Fine (2013) of 39 fMRI studies on sex differences, published in 2009 and 2010, revealed that, for a sample of 22 studies that made only sex comparisons (the other studies also had sex-by-group comparisons which require larger sample sizes), the mean number of males was 13.5 and of females 13.8. Consequently, these studies are especially prone to being underpowered and producing false-positive results.

Closely intertwined with the criticism on the assumption of sex dimorphism in neuroscientific research at the conceptual level is the criticism on how sex/gender is registered in studies involving human participants. Dussauge & Kaiser (2012) criticize the "monolithic character" (p.139) of the sexes/genders studied, as sex/gender is usually assessed by the "simple checking of the F or M box" (Kaiser, 2014).

Concerning data analysis, neurofeminist scholars have pointed out that findings about SGDs can be influenced by the methods used to identify them (Fausto-Sterling, 2000; Kaiser *et al.*, 2009; Bluhm, 2012; Dussauge & Kaiser, 2012; Fine, 2013). For example, Anne Fausto-Sterling (2000) analyzed the extensive body of research on SGDs in the corpus callosum and discovered that whether or not scientists found an SGD seemed to be affected by the methods used. For instance, the way scientists measured the corpus callosum was inconsistent between studies and produced different results. Furthermore, Kaiser et al. (2009) point out that the employed threshold for statistical significance might determine whether an SGD in language processing can be found. In one of their own studies, Kaiser et al. (2007) found that both men and women exhibited a lateral activation of an area associated with language processing at a threshold of $p<0.05$ (Bonferroni [Bonf.] corrected), while at $p<0.01$ (Bonf. uncorrected), men showed a bilateral activation and women did not. As some researchers employ uncorrected p-levels in their analysis of SGDs (Piefke *et al.*, 2005; Clements *et al.*, 2006; Chen *et al.*, 2007) while others use corrected ones (Haller *et al.*, 2005, 2007), Kaiser et al. (2009) stress the importance of uniform conventions to avoid a possible skewing of results.

Besides criticizing inconsistencies in the use of methods, feminist scholars have also argued that some neuroscientific researchers use certain methods in misleading ways or draw impermissible conclusions from them. One instance of this is the critique of the use of and focus on within-group analyses instead of between-group analyses (Bluhm, 2013a; Fine, 2013; Rippon *et al.*, 2014; Hoffman & Bluhm, 2016). Specifically, neurofeminists claim that some researchers focus only on the results of within-group analyses in their discussion, despite performing both kinds of analyses (Shirao *et al.*, 2005; e.g. Hofer *et al.*, 2006). Or that, in less frequent cases, researchers do not even conduct or report on between-group analyses (e.g. Lee *et al.*, 2002). Within-group analyses compare the BOLD signal strength of a particular brain area X during the task of interest with the BOLD signal strength of X during a control task for men and women separately. The outcome of such a within-group analysis could, for example, be that women show significantly higher activity in X during the task than during the control condition while men do not. Neurofeminist scholars claim that some researchers conclude that this demonstrates a neurofunctional SGD for this specific task. However, Bluhm (2013a) points out: "Even if there is an average difference in activity between the groups, if there is also a lot of variability within the groups […], the difference between the groups might not be statistically significant." (p. 323). As a result, neurofeminist scholars criticize the use of within-group analyses to demonstrate differences *between* groups, such as SGDs.

Regarding the interpretation of data analyses, neurofeminist scholars state that there is a tendency in neuroscientific research on SGDs to insufficiently formulate precise hypothesis about the expected differences to be found before the experiment is run. This, as neurofeminists argue, leaves room for the problematic practice of 'HARKing' ('hypothesizing after results are known') which may facilitate the use of stereotypes (Bluhm, 2012, 2013a; Fine *et al.*, 2013). In an analysis of 39 fMRI studies on SGDs, Fine (2013) categorized almost half of them as being either exploratory, so making no prior predictions, or as making only vague predictions. As Fine (2013) criticizes, this practice of not formulating precise a priori hypotheses leave room for "untested, stereotype-infused speculations about the functional significance of neurological findings" (p. 380). This demonstrates the need to develop and test complex neuro-cognitive models that account for the brain areas involved in a specific mental process and the connections between these areas (Bluhm, 2013b, 2013a; Fine, 2013).

*Research Level IIb: Presentation of data and results*

A widespread critique by feminist scholars is that research results on SGDs tend to be presented in an overconfident, exaggerated manner. To be specific, Kaiser et al. (2009) argue that SGDs are emphasized and generalized. For example, they outline a study by Shaywitz et al. (1995) that found an SGD in one aspect of language processing (phonological processing) but no differences in the two other aspects measured. However, in the studies' title, this finding is presented "as evidence for the sex/gender related organization of the brain for *language*" (Kaiser *et al.*, 2009) in general. Fine (2013) agrees that "the confidence with which [findings about differences] are presented by some experts seems premature" (p. 375). Additionally, some feminist scholars have called attention to the citation practices in research on SGDs: For instance, Fine (2013) analyzed 75 papers that cite the Shaywitz et al. (1995) study mentioned above. Even though two meta-analyses have found that, overall, no reliable SGDs seem to exist (Sommer *et al.*, 2004, 2008), more than half of the examined papers did not cite any counterevidence to the claim that there are SGDs in language processing, as voiced by Shaywitz et al. (1995). From the remaining studies, only about a third cited Sommer et al. (2004, 2008) in a fully informative way, meaning the authors mentioned that they are meta-analyses and therefore resume an epistemic advantage over any single study (Fine, 2013). Neurofeminist scholars voice the concern that these kinds of incorrect, partial, or selective citation practices might further the impression that the SGDs described are empirically more supported than they actually are (Kaiser *et al.*, 2009; Fine, 2013).

However, the problem might already start before citation practices can skew the picture of the evidence in a particular direction, as it is likely that fewer studies on sex/gender similarities get published. This publication bias may be based on a general bias in science publishing not to publish any null results, meaning studies reporting to have found no differences between groups (Hoffman & Bluhm, 2016). Nevertheless, feminist scholars argue that this bias is especially problematic for studies about SGDs as it accentuates the sexist biases already present in research (Kaiser *et al.*, 2009; Dussauge & Kaiser, 2012; Meynell, 2012; Fine, 2013).

Lastly, feminist scholars have criticized the pictorial presentations of neuroimaging findings. Meynell (2012) claims that fMRI images often obscure how much groups overlap and how much variance the data contain. Similarly, Kaiser et al. (2009) point out that "dissimilar images quickly lead to the assumption of difference" (p. 54), even though there might not be a difference, which can be the case for findings produced by within-group analyses. To many people, these images appear as "the most natural, immediate, and intuitive method" (Fitsch, 2014) of producing knowledge about the brain. According to Fitsch (2014), this neuroimagery causes an objectification and essentialization of knowledge about the brain and a "re-implementation of gendered stereotypes in brain images" (p.

102).

*Research Level III: Criticism on the level of theory building*

Much criticism from feminist scholars has been directed towards the "'hardwiring' paradigm" (Jordan-Young & Rumiati, 2012). This paradigm describes the idea that sex-differentiating prenatal hormones, especially testosterone, organize SGDs in brain structure and function in a fixed and permanent way. Or, as Fine et al. (2013) put it, it is the idea "that there is a unidirectional, causal pathway from genes to behavior via hormones and brains" (p. 550). Accordingly, 'hardwired' means that researchers regard SGDs as either permanent, innate, meaning genetically determined, or both (Hoffman, 2012). This paradigm is also known as the "brain organization theory" (Fine *et al.*, 2013), which was first proposed by Phoenix et al. (1959) to account for rodent sexual differentiation and reproductive behavior and was later extrapolated to humans (Grossi & Fine, 2012). According to Fine et al. (2013), the brain organization theory has been the dominant doctrine in explaining sexual differentiation of the brain. It has been the backdrop for many theories about SGDs in behavior, such as the influential "Empathizing/Systemizing (E/S) hypothesis" (Grossi & Fine, 2012) by Simon Baron-Cohen (2003).

Many feminist scholars argue that the hardwiring paradigm is unscientific as well as socially, politically, and ethically problematic (Dussauge & Kaiser, 2012; Grossi & Fine, 2012; Hoffman, 2012; Jordan-Young & Rumiati, 2012; Fine, 2013; Fine *et al.*, 2013). The evidence for the hardwiring paradigm has been criticized on many levels: Firstly, feminist scholars have critically assessed the evidence from animal, as well as human studies that aim to support the brain organization theory. Secondly, they have uncovered theoretical flaws in the hardwiring paradigm, and thirdly, they draw attention to the plasticity of the brain and the influence of the environment on it.

As mentioned above, the brain organization theory was developed in the context of rodent research and then extrapolated to human sexual differentiation. Accordingly, research findings on rats and other animals were used as support for the brain organization theory applying to humans. For example, Baron-Cohen (2003) references studies on maze performance in rats and toy preferences in monkeys as evidence for his E/S hypothesis. However, according to Grossi & Fine (2012), many researchers have argued that it is unlikely that rodent and human sexual differentiation are substantially similar and that it is therefore not possible to extrapolate from rats to humans. Additional to animal studies, efforts have been made to prove the brain organization theory in human studies. Since it is obviously unethical to expose human fetuses to certain levels of a specific hormone to study its influence, research has focused on the development of individuals that were naturally exposed to unusual levels of hormones prenatally in cohort studies (Jordan-Young & Rumiati, 2012). According to Fine et al. (2013), these studies often found no support for any "causal links between high absolute levels of testosterone and masculine characteristics" (p. 550). Similarly, Jordan-Young & Rumiati (2012) argue that the data produced by these studies does not support the hardwiring paradigm. For instance, no behavioral differences were found in women who were exposed to high levels of the 'masculinizing' hormone diethylstilbestrol in utero.

Dussauge & Kaiser (2012) also argue that the innatism that underlies the hardwiring paradigm cannot explain the "differentiation and changing character of gender […] in individuals and populations" (p. 136). Moreover, the authors find innatist claims weak unless scientists explicitly uncover the mechanism by which a certain, innate structure causes a specific gendered behavior.

As an alternative paradigm to hardwiring neurofeminist scholars have emphasized the importance of findings about neuronal plasticity. Neuronal plasticity describes the observation that the brain's architecture is malleable structurally and functionally (Hoffman, 2012; Fine, 2013; Fine *et al.*, 2013). This plasticity on the neural and synaptic level is the neuronal basis for learning and memory in the central nervous system (Schmitz & Höppner, 2014a). Therefore, rather than being determined before birth and remaining fixed, the organization of the brain can more accurately be described as a "continuous and dynamic process that persists throughout one's life" (Fine *et al.*, 2013). Accordingly, the concept of plasticity is irreconcilable with the idea that any SGDs are permanent, as well as completely innate since plasticity accounts for an influence of the environment on the brain. As a result, several feminist scholars argue that any SGDs in brain function might actually be due to the gendered environment in which the examined individuals live or to an interplay between environment and genes, rather than solely to prenatal hormones (Grossi & Fine, 2012; Hoffman, 2012; Jordan-Young & Rumiati, 2012; Fine, 2013; Fine *et al.*, 2013). The concept of plasticity is in accordance with and can account for behavioral findings such as sex/gender-typed behaviors being "dramatically modified or even reversed by simple and relatively short-term behavioral interventions such as neonatal handling" (Jordan-Young & Rumiati, 2012). Even abilities that were believed to show robust SGDs, such as mental rotation ability, were demonstrated to be eliminated or even reversed through training (Jordan-Young & Rumiati, 2012). Despite the plasticity paradigm's wide acceptance in the neurosciences (Magee & Grienberger, 2020), many feminist authors argue that the methodology and the way that findings are presented in research on SGDs are rather in line with the hardwiring paradigm.

Apart from the scientific shortcomings of the hardwiring paradigm that feminist scholars have pointed out, some of them have also voiced concerns about the ethical implications of viewing SGDs in the brain as fixed and permanent. These scholars worry that research performed along the lines of the hardwiring paradigm will negatively influence the public perception of gender and gender stereotypes (Fine, 2012, 2013; Jordan-Young & Rumiati, 2012; Fine *et al.*, 2013).

**Discussion**

*The emergence of 'new neurofeminism'*

As we chose the review by Schmitz & Höppner (2014b) as a starting point for our analysis, we especially focus on the neurofeminist literature that was published after the review to reveal more recent trends of the neurofeminist discourse. When outlining the results of our analysis of the 'newer' neurofeminism, we mirror the structure we used when outlining the more detailed criticism on the three levels of focus. However, we restrict our findings to the conceptual and methodological level, as the discussions and developments on the level of theory building are so diverse and far-reaching that they go beyond the scope of this paper.

*Conceptual considerations: Sex/gender as a variable*

At the conceptual level, a clear trend towards developing and refining arguments against the idea that brains are sexually dimorphic can be noted. In this context, two concepts from gender scholarship have received increased attention in newer neurofeminist discourses: overlap and mosaicism. Multiple papers refer to these key principles to argue against sex/gender dimorphism and essentialist notions of the relationship between sex/gender and the brain. Additionally, scholars use these concepts as a starting point for developing frameworks on how to include the variable sex/gender in neuroscientific research in a more sensitive, neurofeminist-informed way.

Bluhm (2012) argues that sex/gender should be treated as a "complex, multilevel, hierarchical structure" (p. 551) in research. The first concrete proposition to register sex/gender in a more complex, multiparametric way comes from Kaiser (2014). She proposes seven different sex/gender markers, social as well as biological ones, to segregate and register many different aspects of what is "simplistically called 'sex' or 'gender' in experimental settings" (Kaiser, 2014). Furthermore, Kaiser (2014) develops ideas on how to operationalize a sexed/gendered variable in an experiment in a non-essentialized way, based on two exemplary conditions. These ideas are, as the author herself points out, "only first thoughts" (p. 57) for accounting for a more complex, unfixed, and changeable definition of sex/gender. In her paper, she defines some next working steps for explicating the multiparametric registration of gender further.

The concept of overlap, as already touched upon above, describes the idea that the distribution of 'male' and 'female' brain phenotypes is highly overlapping. Most feminist scholars now refer to this idea with the term "mosaicism" (e.g. Rippon *et al.*, 2014), as they believe the human brain to be a mosaic of "male-typical and female-typical features" (Fine, Joel, *et al.*, 2019, n.p.). According to Rippon et al. (2014), neuroscientists have recognized the concept of brain mosaicism for decades. However, many feminist scholars find the application of this concept to neuroscientific research on SGDs unsatisfying or lacking (Rippon *et al.*, 2014; Hoffman & Bluhm, 2016; Joel & Fausto-Sterling, 2016; Bentley *et al.*, 2019a; Bryant *et al.*, 2019; Hyde *et al.*, 2019; Jordan-Young *et al.*, 2019; Eliot *et al.*, 2021). As a result, feminist scholars have increasingly engaged with the phenomenon of mosaicism, and in this context often also with the concept of overlap, so that these concepts arrive in the mainstream neuroscience research on SGDs. Firstly, feminist neuroscientists conducted empirical research to investigate whether the findings from animal studies actually apply to the human brains as hypothesized and if so, how common mosaicism and overlap in the brain are (Joel *et al.*, 2015; Hyde *et al.*, 2019; Alon *et al.*, 2020; Eliot *et al.*, 2021). Secondly, neurofeminists worked out the implications these empirical findings have for future research on SGDs (Rippon *et al.*, 2014, 2017; Joel & Fausto-Sterling, 2016; Bentley *et al.*, 2019b; Hyde *et al.*, 2019; Alon *et al.*, 2020).

As mentioned above, Joel (2014) backed up her initial arguments against the internal consistency of brains with findings from animal studies only. In order to advance her claims, Joel et al. (2015) conducted a study to evaluate the degree of internal consistency of structural features in human brains, using MRI data of more than 1400 brains from four different data sets. The data were retrieved from several different imaging techniques and the authors analyzed them with various methods, ensuring that the outcomes were generalizable across imaging techniques and methods of analysis. A similar study was conducted by Alon et al. (2020). The authors analyzed 23,935 brain scans of females and males in different MRI-derived measures to test the masculinization hypothesis. This theory assumes that "specific features in the brains of males are masculinized away from a default female form" (Alon *et al.*, 2020) by sex-related factors, forming two distinct, sexually dimorphic population of brains. Furthermore, two syntheses on the evidence for brain mosaicism exist (Hyde *et al.*, 2019; Eliot *et al.*, 2021).

Firstly, Hyde et al. (2019) synthesized research that challenges the sex/gender binary from five different fields. Secondly, Eliot et al. (2021) summarized the evidence for SGDs in many areas of brain research, namely structural and functional SGDs, differences in lateralization and interhemispheric connectivity, and connectome differences.

The outcome of this empirical work underlines previous claims about overlap and mosaicism of SGDs in the brain. Firstly, the research highlights that structural as well as functional SGDs in the human brain are often statistically small and that there is a substantial overlap (Rippon *et al.*, 2014; Joel & Fausto-Sterling, 2016; Alon *et al.*, 2020; Eliot *et al.*, 2021). As a result, sex/gender is found to assume only a minor role in explaining the variability in human brain structure and function (Alon *et al.*, 2020; Eliot *et al.*, 2021). Secondly, these studies have provided substantial evidence against the idea that brain features are internally consistent within individual human brains and therefore for the concept of brain mosaicism in human brains (Joel *et al.*, 2015; Hyde *et al.*, 2019; Alon *et al.*, 2020; Eliot *et al.*, 2021).

According to Bryant et al. (2019), the mosaic model has "greater explanatory power for understanding the relationship between sex/gender and brain organization" than a dichotomous understanding of sex/gender (Rippon *et al.*, 2014; see also Bentley *et al.*, 2019a; Hyde *et al.*, 2019). This criticism of defining sex/gender in the traditional, dichotomous way has already been present in the first wave of neurofeminist scholarship and was discussed above. As a result, the more recent neurofeminist discussions have increasingly focused on the question of how to move away from a dichotomous framework of comparing men and women, and how to operationalize the variable sex/gender in neuroscientific research in a more informative way (Rippon *et al.*, 2014, 2017; Joel & Fausto-Sterling, 2016; Bentley *et al.*, 2019b; Bryant *et al.*, 2019; Gungor *et al.*, 2019; Hyde *et al.*, 2019; Shattuck-Heidorn & Richardson, 2019). In the following, we summarize the most common considerations and recommendations proposed by the 'newer' neurofeminism.

*Recommendations for neuroscience research from neurofeminist scholarship*

According to Bryant et al. (2019), there is no "solid alternative measurement to comparing women and men" yet. Other neurofeminist authors seem to agree with this statement as they call for further research on this matter (Joel & Fausto-Sterling, 2016; Bentley *et al.*, 2019b; Alon *et al.*, 2020). For instance, Joel & Fausto-Sterling (2016) request more research on the relationship between sex/gender and the brain and on the question of when to include the sex/gender category as a variable and when not to. Until more information about these questions exists, some authors outline situations in which the sex/gender category should be included. For instance, Bentley et al. (2019b) suggest that using a binary sex/gender category might be useful to conduct "bridge projects" and follow up on existing research. Furthermore, Joel & Fausto-Sterling (2016) regard the sex/gender variable valuable for the study of brain pathologies that show a different prevalence in males and females. At the same time, some authors advocate for replacing the sex/gender category by social and psychological variables that might correlate with sex/gender (Joel & Fausto-Sterling, 2016; Hyde *et al.*, 2019) or supplementing it with these variables (Rippon *et al.*, 2017; Bentley *et al.*, 2019b; Hyde *et al.*, 2019). Additionally, some authors suggest that demographic data about the test subjects should be collected to reflect the "entangled complexity of their psychological, physical, and material experiences" (Rippon *et al.*, 2014; see also Bentley *et al.*, 2019b). Furthermore, Bentley et al. (2019b) and Rippon et al. (2014) recommend exploring the context in which sex/gender differences appear and disappear. This practice would enable a deeper analysis of a purported relationship between sex/gender and brain function, for example by being able to investigate whether this relationship is mediated or moderated by (an)other variable(s) (Rippon *et al.*, 2017).

Lastly, feminist scholars also discuss how gender can be registered in a neuroscientific experiment. Some authors stress the multidimensionality of the variable, meaning that gender is comprised of different levels, such as the levels of societal norms, structural forces, and patterns, and sets of internal beliefs (Shattuck-Heidorn & Richardson, 2019; see also Bentley *et al.*, 2019a; Hyde *et al.*, 2019). They propose a "multiparametric registration of sex/gender" (Rippon *et al.*, 2014), similar to the approach by Kaiser (2014) mentioned above. The authors suggest that gendered personality dimensions, gender attitudes and self-attributed gender norms should be assessed, for example through the use of questionnaires (Rippon *et al.*, 2014; Bentley *et al.*, 2019b). Furthermore, Hyde et al. (2019) advocate for the conceptualization and measurement of gender in nonbinary ways. To implement this, researchers could provide more options (e.g., genderqueer), ask open-ended questions and use continuous measures (Hyde *et al.*, 2019).

As described above, early neurofeminism devoted great effort towards critically assessing the underlying assumptions of neuroscientific SGD research and the ways in which sexist biases can enter this research. While neurofeminists precisely outlined the different methodological flaws that distorted the evidence for SGDs in the brain, they were reserved in proposing other, better ways to capture the relationship between sex/gender and the brain. Of course, one might deduce from the earlier criticism how or what *not* to do, and proposals for better research designs and presentation of results were already beginning to emerge before 2014. However, the dedication to developing and applying feminist interventions rather than just criticizing the status quo in the newer neurofeminist literature sets it apart from older neurofeminism.

An important part of this endeavor is to develop, adapt and outline "feminist tools" (Bryant *et al.*, 2019, n.p.) that are supposed to ensure that SGD research is conducted in "the most informative and useful way" (ibid.). One quite popular suggestion in newer neurofeminist writings is the pre-registration of protocols (Rippon *et al.*, 2014, 2017, 2021; Hoffman & Bluhm, 2016; Bentley *et al.*, 2019b). This practice requires that a research protocol, with all the details of a planned study, such as which comparisons are going to be made and what the hypotheses are, is submitted and peer-reviewed in advance. This way, reviewers can comment on the research design before it is carried out and draw attention to possible methodological flaws. Additionally, this practice addresses the problem of publication bias since researchers must report the outcomes of all registered analyses. Also, the pre-registration of research protocols restrains the practice of reverse inferences, i.e. hypothesizing after results are known ('HARKing'). However, the practice of pre-registration does not address all methodological issues raised by feminist scholars, such as whether results are presented in an essentialist framework (Bentley *et al.*, 2019b). Accordingly, additional "feminist tools" (Bryant *et al.*, 2019, n.p.) are necessary.

Other common requests by feminist scholars are the inclusion of effect sizes (Rippon *et al.*, 2014, 2017; Joel & Fausto-Sterling, 2016; Bryant *et al.*, 2019) and other measures of overlap, such as the "Index of Similarity" (Rippon *et al.*, 2017; see also Bryant *et al.*, 2019; Rippon *et al.*, 2021). In general, feminist scholars call for a greater focus on similarity. Some authors, for instance, request an "institutionalization of sex/gender similarity" (Rippon *et al.*, 2014) in databases (see also Hoffman & Bluhm, 2016). Lastly, several different scholars have proposed an array of different statistical analyses that go beyond standard parametric tests and, for example, allow for the inclusion of several variables and the measurement of their interrelationship (Rippon *et al.*, 2014, 2017; Bentley *et al.*, 2019b; Bryant *et al.*, 2019; Hyde *et al.*, 2019).

As an outcome of these considerations on how to make the research on SGDs more feminist, some authors have developed guidelines or templates for future research (Rippon *et al.*, 2014, 2017; Joel & Fausto-Sterling, 2016; Bentley *et al.*, 2019b; Bryant *et al.*, 2019). For instance, Bentley et al. (2019b) developed an "advisory framework" for the research on spatial cognition. A further example is the set of guidelines developed by Rippon et al. (2014) to ensure that neuroimaging research on SGDs has addressed key principles from sex/gender scholarship.

Bryant et al. (2019) went one step further and implemented many of their proposed feminist methodological interventions in a case study.

*Reaching out: The reception of neurofeminist works and neurofeminist engagement*

A second theme that stand out in more recent neurofeminist scholarship is the increased focus on disseminating neurofeminist scholarship on SGD research and engage in bidirectional communication with the public and stakeholders in neuroscience. For example, some neurofeminist articles have been translated into Spanish, making neurofeminist debates accessible to a broader audience (Bovet *et al.*, 2013; Kaiser, 2018; Fine, Jordan-Young, *et al.*, 2019). Furthermore, one overview by Fine, Joel et al. (2019) is an educational text on how to interpret putative SGDs obtained by neuroscientific research, written explicitly for people with no background in neurosciences. Neurofeminist scholars have also reached out to their colleagues from the neurosciences to move the debate about a more critical, gender-sensitive neuroscientific research on SGDs into 'mainstream' neuroscience discourses (Rippon *et al.*, 2017; Gungor *et al.*, 2019).

*Outlook for further research*

From our perspective, it would be of great interest to further investigate how neurofeminist critiques were received in the mainstream neurosciences in more detail. To this regard, some of the following questions could be explored: Are the 'mainstream' neurosciences aware of these critiques? If yes, do they respond to them or ignore them? If they answer, what is the content of these answers? What are points of agreement and disagreement? As this paper has only evaluated one side of the debate, it is important to also look at other perspectives. This could help to interpret neuroscientific findings of SGDs in a more informed way, which would be a vital next step for advancing neurofeminist scholarship and its dissemination in the neurosciences.

**Statement regarding potential conflicts of interest**

The authors have no conflict of interest to declare.

**Author contributions**

Kassandra Friedrichs: literature research, first draft, editing

Philipp Kellmeyer: conceptualization, research design, co-writing, editing

**Data accessibility**

The paper provides a narrative review, no quantitative data were gathered for the study.

**List of abbreviations**

| | |
|---|---|
| 3G | genetic, gonadal, genital |
| Bonf. | Bonferroni |
| E/S hypothesis | empathizing/systemizing hypothesis |
| fMRI | functional magnetic resonance imaging |
| HARKing | hypothesizing after results are known |
| SGD | sex/gender differences |
| p. | page |
| PET | positron emission imaging |